\documentclass[%
 aip,
 amsmath,amssymb,
 reprint,%
]{revtex4-1}

\usepackage{graphicx}
\usepackage{dcolumn}
\usepackage{bm}

\usepackage[utf8]{inputenc}
\usepackage[T1]{fontenc}
\usepackage{mathptmx}
\usepackage{etoolbox}
\usepackage{setspace}

\usepackage{lineno}

\usepackage{hyperref}

\makeatletter
\def\@email#1#2{%
 \endgroup
 \patchcmd{\titleblock@produce}
  {\frontmatter@RRAPformat}
  {\frontmatter@RRAPformat{\produce@RRAP{*#1\href{mailto:#2}{#2}}}\frontmatter@RRAPformat}
  {}{}
}%
\makeatother
\begin{document}

\preprint{AIP/123-QED}

\title[]{Elevating electron energy gain and betatron X-ray emission in proton-driven wakefield acceleration}
\author{Hossein Saberi}
\author{Guoxing Xia}%

 \author{Linbo Liang}
\affiliation{ 
Department of Physics and Astronomy, University of Manchester, Manchester M13 9PL, United Kingdom
}%
\affiliation{%
	Cockcroft Institute, Daresbury, Cheshire WA4 4AD, United Kingdom
}%

\author{John Patrick Farmer}

\affiliation{CERN, 1211 Geneva, Switzerland}
\affiliation{%
Max Planck Institute for Physics, 80805 Munich, Germany
}%

\author{Alexander Pukhov}
\affiliation{Heinrich-Heine-Universität Düsseldorf, 40225 Düsseldorf, Germany}

\email{guoxing.xia@manchester.ac.uk}
\email{hossein.saberi@manchester.ac.uk.}

\date{\today}
\begin{abstract}
The long proton beams present at CERN have the potential to evolve into a train of microbunches through the self-modulation instability process. The resonant wakefield generated by a periodic train of proton microbunches can establish a high acceleration field within the plasma, facilitating electron acceleration. This paper investigates the impact of plasma density on resonant wakefield excitation, thus influencing acceleration of a witness electron bunch and its corresponding betatron radiation within the wakefield. Various scenarios involving different plasma densities are explored through particle-in-cell simulations. The peak wakefield in each scenario is calculated by considering a long pre-modulated proton driver with a fixed peak current. Subsequently, the study delves into the witness beam acceleration in the wakefield and its radiation emission. Elevated plasma density increases both the number of microbunches and the accelerating gradient of each microbunch, consequently resulting in heightened resonant wakefield. Nevertheless, the scaling is disrupted by the saturation of the resonant wakefield due to the nonlinearities. The simulation results reveal that at high plasma densities an intense and broadband radiation spectrum extending into the domain of the hard X-rays and gamma rays is generated. Furthermore, in such instances, the energy gain of the witness beam is significantly enhanced. The impact of wakefield on the witness energy gain and the corresponding radiation spectrum is clearly evident at extremely elevated densities.
\end{abstract}

\maketitle
\section{\label{sec:level1}Introduction}
Electrons can be efficiently focused and accelerated within the plasma wakefield generated by a high-intensity driver, e.g., laser pulse,~\cite{Esarey2009} electron-~\cite{Hogan2016} or a proton beam,~\cite{Adli2016} due to the presence of very strong transverse and longitudinal electric wakefields. The wiggling electrons then emit electromagnetic radiation referred to as betatron radiation (BR)~\cite{Esarey2002,Kostyukov2003,PhysRevLett.88.135004} that is usually characterized by a high brightness, synchrotron-like broadband radiation. The pulse duration is equivalent to that of the electron bunch, typically on the femtosecond scale, in various wakefield accelerator concepts. BR in laser-plasma accelerators has been observed experimentally in the keV photon energy range,~\cite{Rousse2004} pursued by advanced experiments to boost the peak brightness~\cite{Kneip2010} and expand the energy range.~\cite{Cipiccia2011} 
Advanced single-shot phase contrast imaging with BR has been demonstrated in experiment.~\cite{Fourmaux2011} 
Furthermore, BR can serve as a non-destructive diagnostic tool for assessing electron beam parameters. As a result, betatron spectroscopy has attracted attention in various wakefield accelerator contexts.~\cite{Kneip2012,Kohler2016,Curcio2017,Shpakov2016,Claveria2019,Williamson2020,Liang2023}

In contrast to laser pulses and electron beams, the proton driver maintains its energy in a plasma medium over much longer distances ranging from hundreds to thousands of metres.~\cite{Caldwell2011} The Advanced Wakefield Experiment (AWAKE) Run 1 (2016-2018) at CERN has successfully demonstrated the proton-driven plasma wakefield acceleration (PD-PWFA) using the Super Proton Synchrotron (SPS) proton beam.~\cite{Adli2018} Currently, the AWAKE Run 2 (2021-) is focused on achieving the generation of a multi-GeV electron beam while controlling its emittance, which holds potential for particle physics experiments.~\cite{Gschwendtner2022} To mitigate the beam filamentation instability, AWAKE deliberately operates at a density where the plasma skin depth is on the order of the radius of the SPS proton driver. The acceleration of the witness beam and the corresponding BR at this plasma density have been investigated in previous studies.~\cite{Williamson2020,Liang2023} These studies demonstrated that the electron beam accelerates up to several GeV energies. Simultaneously, the betatron spectrum was observed in the UV to X-ray range. 

Opting for the higher plasma density would necessitate a smaller proton driver radius. Research into proton-driven wakefield at high densities is less extensive compared to that conducted in the AWAKE. In a simulation study utilizing the long proton driver proposed for the RHIC-EIC project at the Brookhaven National Lab, proton beams with radii of $100 \, \mu \mathrm{m}$ and $40 \, \mu \mathrm{m}$ have been considered.~\cite{Chappell2019} The results revealed a significant increase in the peak wakefield.

In this study, particle-in-cell (PIC) simulations are conducted to explore the influence of plasma density on the general case of resonantly driven wakefield acceleration. The peak wakefield in each scenario is determined by considering a long pre-modulated proton driver with a fixed peak current. The strong scaling observed at low densities is ultimately constrained by saturation of the wakefield growth at extremely high densities. Subsequently, the investigation centers on electron acceleration within the peak wakefield, while also examining the corresponding radiation emission. The paper is structured as follows: Section~\ref{sec3} provides an overview of proton-driven wakefield. In Sec.~\ref{sec4}, the simulation model and the methodology are elucidated in detail. In Sec.~\ref{Sec4-1}, the results are presented and discussed. Conclusions are summarized in Sec.~\ref{sec5}.
\section{\label{sec3}Proton-driven wakefield}
The PD-PWFA concept, first proposed by Caldwell et al,~\cite{Caldwell2009} has introduced a pioneering approach to future plasma-based colliders that could be realized within a single acceleration stage.~\cite{Xia2014} While AWAKE Run 1 has marked significant strides in advancing the realization of PD-PWFA using SPS proton beam,~\cite{Adli2018} AWAKE Run 2 aims to achieve a more stable beam suitable for applications in particle physics experiments.~\cite{Gschwendtner2022}
\begin{figure}[]
	\centering
	\includegraphics[angle=0, width=.49\textwidth]{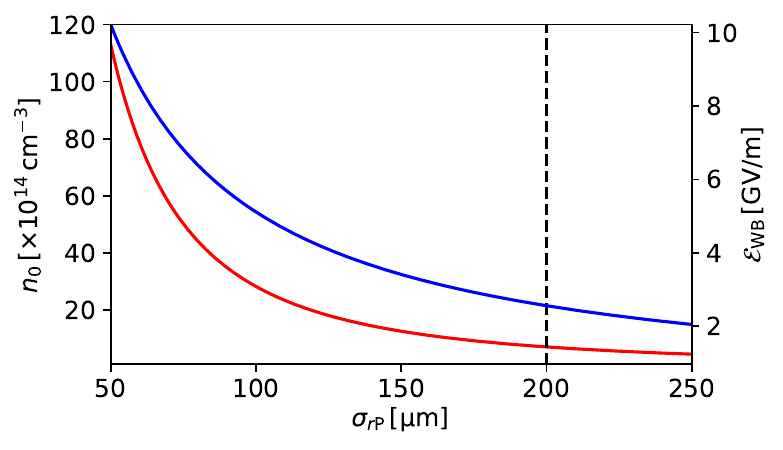}
	\caption{\footnotesize{Maximum plasma density (red line) and the wavebreaking limit (blue line) versus the proton beam radius, ensuring prevention of beam filamentation instability. The dashed line indicates the SPS beam radius.}}
	\label{single}
 \label{proton}
\end{figure} 
The geometry of the SPS beam imposes limitations on its suitability for plasma wakefield excitation. The SPS beam has a few centimeter-scale length, while it should be on the order of the plasma wavelength, i.e., sub-millimeter scale, to effectively generate wakefields in the plasma.  However, this obstacle is overcome by an intrinsic plasma instability. The long proton beam undergoes self-modulation instability (SMI) in the plasma, resulting in the formation of a train of microbunches with a period on the order of the plasma wavelength.~\cite{Kumar2010} Although SMI is a complex phenomenon, it has undergone comprehensive examination and validation through extensive studies conducted in the AWAKE experiment.~\cite{Marlene2019}

The current filamentation instability can destroy the proton beam when interacting with the plasma. To avert this instability, the proton beam radius, $\sigma_{r\mathrm{P}}$, needs to be on the order of or smaller than the plasma skin depth~\cite{Allen2012}, or 
\begin{equation}
k_\mathrm{p} \sigma_{r\mathrm{P}} \leq 1,
\label{matchpr}
\end{equation}
where $k_\mathrm{p} =2 \pi /\lambda_\mathrm{p}$ and $\lambda_\mathrm{p}$ is the plasma wavelength. The normalized emittance of the SPS proton beam is 3 mm mrad, and so the beam is focused to a radius of 200 $\mu\mathrm{m}$. In this case, a maximum plasma density of $7 \times 10^{14} \,\mathrm{cm^{-3}}$ is required to meet the condition outlined in Eq.~(\ref{matchpr}).

Plasma density is pivotal in wakefield acceleration as it directly influences the wavebreaking limitation. Nevertheless, opting a higher plasma density would require proton driver with smaller radius as indicated by Eq. (~\ref{matchpr}). For instance, a proton driver with a radius of $100 \, \mathrm{\mu m}$ could be used with the plasma density of $2.8 \times 10^{15} \, \mathrm{cm^{-3}}$. The maximum plasma density and the wavebreaking limit, $\mathcal{E}_\mathrm{WB}$, versus the proton beam's radius are illustrated in Fig.~\ref{proton}.

In the following section, we delve into the impact of plasma density on the wakefield generated by a long pre-modulated proton driver and determine its peak magnitude. Additionally, we investigate electron acceleration within the peak wakefield using PIC simulations. The three-dimensional quasi-static PIC code QV3D, built upon the VLPL code platform,~\cite{pukhov_1999} is employed. The code incorporates a built-in module to calculate the synchrotron radiation for each macroparticle.
\begin{table*}
\caption{\label{plasmaparam} Plasma parameters for comparative simulation study of a Baseline scenario and three distinct cases with higher plasma densities.}
\begin{ruledtabular}
\begin{tabular}{cccccc}
 Plasma parameter & Symbol [Unit] & Baseline &  Case I & Case II & Case III \\ \hline
\textbf{}&&&&&\\
Skin depth & $[\mathrm{\mu m}]$ &  200 & 150 & 100 & 50 \\
 Density &$n_0\, [\mathrm{cm^{-3}}]$& $7\times10^{14}$ & $1.25 \times 10^{15}$ & $2.80 \times 10^{15}$ & $1.13 \times 10^{16}$ \\
 Length & [m] & 10 & 10 & 10 & 10 \\
 \end{tabular}
\end{ruledtabular}
\end{table*}

\section{Simulation configuration}
\label{sec4}
The simulation study investigates witness beam acceleration and its associated radiation emission in a baseline density alongside three specific cases featuring higher plasma densities. The Baseline scenario includes a plasma density of $7\times 10^{14} \, \mathrm{cm^{-3}}$ and a proton driver with a radius of $200\, \mu \mathrm{m}$. The Baseline parameters are selected to yield the same peak wakefield amplitude as in.~\cite{Olsen2018,Liang2023} Three other cases are considered with proton driver radii of $150\, \mu \mathrm{m}$ (Case I), $100\, \mu \mathrm{m}$ (Case II) and $50\, \mu \mathrm{m}$ (Case III). 
Plasma parameters are summarized in Table~\ref{plasmaparam}
 \subsection{Peak wakefield} \label{sec4A}

\begin{table}
\caption{\label{micro} Microbunch parameters in the periodic train driver with a total of $1.62 \times 10^{10}$ protons spanning a length of 32.8 mm.}
\begin{ruledtabular}
\begin{tabular}{ccccc}
  Parameter [Unit] &  Baseline &  Case I & Case II & Case III\\ \hline
\textbf{}&&&&\\
No. Microbunch & 26 & 34 & 52 & 104 \\
Length $[\mathrm{\mu m}]$ &  283 & 213 & 142 & 71 \\
Radius $[\mathrm{\mu m}]$ &   200 & 150 & 100 & 50\\
Charge $[\mathrm{pC}]$ &   100 & 76 & 50 & 25\\
\end{tabular}
\end{ruledtabular}
\end{table}
\begin{table*}
\caption{\label{pealwake} The peak wakefield of the train driver calculated by QV3D simulations, $\mathcal{E}_\mathrm{peak}$, and using the linear theory formulae, the wavebreaking limit $\mathcal{E}_\mathrm{WB}$ and the Ratio parameter.}
\begin{ruledtabular}
\begin{tabular}{ccccc}
  &  $\mathcal{E}_\mathrm{peak}$ [GV/m] & Linear theory value [GV/m]  & $\mathcal{E}_\mathrm{WB}$ [GV/m] & Ratio \\ \hline
\textbf{}&&&\\
Baseline & 0.45  & 0.44  & 2.55 & 18\% \\
Case I & 0.76  & 0.77  &  3.40 & 22 \%\\
Case II & 1.71  & 1.73   & 5.09 & 34 \%\\
Case III & 3.9 $\,$  & 7.00  & 10.22 & 38 \%\\
\end{tabular}
\end{ruledtabular}
\end{table*}
\begin{figure*}[]
	\centering
	\includegraphics[angle=0, width=.99\textwidth]{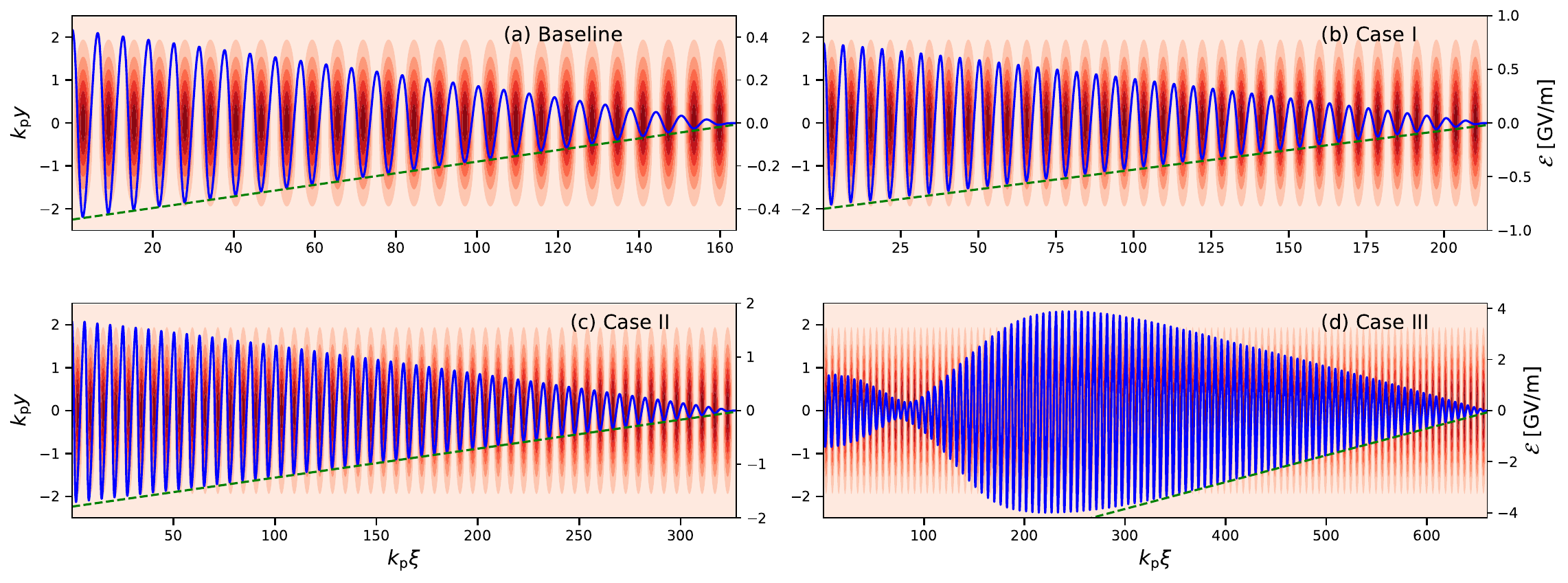}
	\caption{\footnotesize{The wakefield (blue line) of a train of microbunches (red contours) for Baseline (a), Case I (b), Case II (c) and Case III (d). The microbunch parameters are provided in Table~\ref{micro}. The green dashed lines represent the linear superposition of the wakefields.}}
	\label{train}
\label{singleparticle}
\end{figure*}
The self-modulation process within the plasma divides a lengthy proton beam into short microbunches, each spaced by the plasma wavelength. These microbunches resonantly excite a substantial wakefield,~\cite{Kumar2010} a phenomenon demonstrated in both the AWAKE experiment~\cite{Marlene2019} and simulations conducted at higher plasma densities.~\cite{Chappell2019}

In this context, rather than conducting a full PIC simulation of the self-modulation of a long proton beam, we provide an estimation of the peak wakefield through the utilization of a pre-modulated proton driver. For all scenarios, we consider a proton beam comprising $1.62 \times 10^{10}$ particles, spanning a length of 32.8 mm. The beam is structured as a periodic sequence of microbunches, with a period equal to the plasma wavelength. The charge is chosen to ensure that in Baseline the pre-modulated driver generates a peak wakefield of 450 MeV/m.~\cite{Olsen2018,Liang2023} Gaussian microbunches with equal charges are used, facilitating an estimation of the peak wakefield. Adhering to the criterion in Eq.~(\ref{matchpr}), the maximum beam radius is utilized in each scenario. The number of microbunches can be determined through a straightforward algebraic calculation. For instance, in the Baseline scenario, 26 microbunches are present with a spacing equivalent to the plasma wavelength. The length of each microbunch is chosen as 
\begin{equation}
    k_\mathrm{p} \sigma_{\xi\mathrm{mb}} = \sqrt{2},
\end{equation}
to maximize the wakefield.~\cite{Lu2005} Here, $\sigma_{\xi \mathrm{mb}}$ represents the length of each microbunch. The parameters of the microbunches are summarized in Table~\ref{micro}.

A simulation of a periodic train driver is conducted, with each microbunch's parameters remaining consistent with those outlined in Table~\ref{micro}. Figure \ref{train} depicts the periodic train driver and the initial wakefield in various scenarios. The notable rise in the peak wakefield at elevated plasma densities can be attributed to both the augmented number of microbunches and the heightened wakefield of each microbunch. Each microbunch density is much smaller than the plasma density; therefore, the peak wakefield can be estimated using the linear theory formula as~\cite{Lu2005}
  \begin{equation} \label{linear}
\mathcal{E}_\mathrm{mb}[\mathrm{MV/m}]  =  240 \left (\frac{N_\mathrm{mb}}{4 \times 10^{10}} \right ) \left (\frac{600 \, \mu \mathrm{m}}{\sigma_{\xi\mathrm{mb}}} \right )^2,
\end{equation}
where $N_\mathrm{mb}$ is the number of particles in each microbunch. In the linear regime, we can assume a linear superposition of the wakefields to calculate the net wakefield of the modulated proton driver. The green dashed lines in in Fig.~\ref{train}(a)-~\ref{train}(c) make it evident that the wakefields of microbunches superimpose linearly reaching a maximum at the end. However, in Fig.~\ref{train}(d), the wakefield grows up to a certain point and then saturates. This observation aligns with the understanding that the wakefield of a train of microbunches may decay at some point (about 40\% of the wavebreaking field) due to the plasma non-linearities, i.e., the nonlinear elongatoin of the wave period that leads to the de-phasing between the microbunches and the wakefield~\cite{Lotov2014}. In Fig.~\ref{train}(d), the saturated wakefield exhibits a renewed increase due to the linear superposition of the remaining microbunches within the driver. The wakefield saturation could potentially occur in other scenarios if additional microbunches are incorporated into the driver. The same result is evident in Table~\ref{pealwake}, where a comparison of peak wakefields in simulation, $\mathcal{E}_\mathrm{peak}$, and theory is provided. Table~\ref{pealwake} also incorporates information on wavebreaking, $\mathcal{E}_\mathrm{WB}$,  and the Ratio, calculated as the percentage of $\mathcal{E}_\mathrm{peak}$/$\mathcal{E}_\mathrm{WB}$, providing a clearer understanding of the peak wakefield. 

It is noteworthy that in the AWAKE experiment, as the modulated proton driver travels within the plasma, the amplitude of the wakefield decays after reaching saturation. It is proposed that introducing a density step into the plasma helps the wakefields maintain a near-saturation amplitude for an extended distance along the plasma.~\cite{Lotov2020}
\begin{table*}
\caption{\label{AWAKEparam} Parameters for simulation of the witness beam acceleration in the wakefield of the dummy driver for various scenarios.}
\begin{ruledtabular}
\begin{tabular}{cccccc}
  Parameter & Symbol [Unit] & Baseline &  Case I & Case II & Case III \\ \hline
\textbf{}&&&&&\\
 \textbf{Dummy Driver}&&&\\
Energy & $E_{\mathrm{P}}\, \mathrm{[GeV]}$  & 400       & 400 & 400 & 400 \\ 
Charge  & $Q_\mathrm{P}\, [\mathrm{nC}]$ & 2.34 & 2.37 & 2.51 & 2.02 \\
Density & $n_\mathrm{P}/n_0$ & 0.83 & 0.83& 0.88 & 0.71 \\
Bunch Length & $\sigma_{\xi \mathrm{P}}\, [\mathrm{\mu m}]$ & 40 & 40 & 40 & 40\\
Bunch Radius & $\sigma_{r\mathrm{P}}\, [\mathrm{\mu m}]$ &  200 & 150 & 100 & 50 \\\\
\textbf{Electron Witness}&&&\\
Density &$n_\mathrm{e}/n_0$  & 34.1 & 32.7 & 32.3 & 18.1  \\
Charge   &  $Q_\mathrm{e} \, [\mathrm{pC}]$ &  120	& 115 & 113 & 64 \\
Bunch Length & $\sigma_{\xi \mathrm{e}}$ [$\mathrm{\mu m}$] & 60 & 45 & 30 & 15 \\
Bunch Radius & $\sigma_{r\mathrm{e}}\, [\mathrm{\mu m}]$ & 5.75 & 4.98 & 4.07 & 2.87  \\
Energy & $E_{\mathrm{e}} \, [\mathrm{MeV}]$ & 150	 &  150    & 150  & 150 \\ 

Energy Spread &$\delta E_\mathrm{e}$& 0.1\%& 0.1\% & 0.1\% & 0.1\% \\ 
Normalized Emittance & $\varepsilon_{\mathrm{ne}}$ [$\mathrm{mm\, mrad}$] & 2 & 2 & 2& 2\\
\end{tabular}
\end{ruledtabular}
\end{table*}
 \subsection{Witness beam acceleration in peak wakefield} 
We employ a simulation model introduced by Olsen et al~\cite{Olsen2018} to investigate the electron beam acceleration in the peak wakefield of the pre-modulated drivers in Sec~\ref{sec4A}. The simulation model comprises a short proton bunch as the driver and a trailing witness electron beam moving in the background plasma. The proton mass of the driver is multiplied by $10^{6}$ and the emittance is set to zero, to render the proton driver highly rigid regardless of the beam radius, and hence neglect the wakefield variation along the plasma. The dummy driver is selected in such a way that its peak wakefield mirrors that of the pre-modulated driver beam. This model is chosen to reduce the simulation size and, consequently, the computational requirements. 
While the model is highly convenient, it imposes certain physical constraints. Since the evolution of the driver is neglected, it is only appropriate in regimes in which the limiting factor on acceleration is the plasma length, and not dispersion or depletion of the driver.  This would also require the emittance of the initial bunch to be scaled with the driver radius to avoid a stronger divergence for narrow beams.

With the peak wakefield identified in all scenarios, the parameters for the dummy driver in the toy model can now be established. It shares the same radius as the train driver in each scenario. In all cases, the bunch length is assumed to be identical to that in the Baseline scenario which is 40 $\mu \mathrm{m}$. However, the charge is optimized using QV3D simulations to align its peak wakefield with $\mathcal{E}_\mathrm{peak}$ detailed in Table~\ref{pealwake}. It is important to note that the dummy driver should maintain a relatively small density, specifically $n_\mathrm{P}/n_0 \lesssim 1$, to ensure its operation within the quasi-linear regime.~\cite{Rosenzweig} 
 
The witness bunch radius at the entrance of the plasma must be adjusted to ensure that the witness emittance pressure matches the focusing force from the blowout. This adjustment prevents beam radius oscillation and consequently controls emittance growth. For a Gaussian electron beam, the matched radius is determined by~\cite{Olsen2018}  
\begin{equation}
    \sigma_{r\mathrm{e}}=\left(2\epsilon_\mathrm{ne}^2/ \gamma_\mathrm{e}k_\mathrm{p}^2\right)^{1/4},
    \label{ematch}
\end{equation}
where $\epsilon_\mathrm{ne}$ and $\gamma_\mathrm{e}$ represent the normalized emittance and the Lorentz factor of the witness beam, respectively. This results in distinct beam radii for each scenario. The beam length should be less than $\lambda_\mathrm{p}/4$ to ensure that the witness beam remains in the accelerating phase of the wakefield. As a result, $k_\mathrm{p}\sigma_{\xi\mathrm{e}}$ is chosen to be constant in all scenarios. This leads to a shorter witness lengths at higher plasma densities. The witness charge is selected such that $I_\mathrm{e} = 1.33 \times 10^3 \mathcal{E}_\mathrm{peak}/\mathcal{E}_\mathrm{WB}$ to maintain consistent beam loading across different scenarios. This configuration ensures that the entire bunch experiences an almost identical electric field, leading to collective acceleration and the production of a quasi-monoenergetic beam with minimal energy spread. However, it results to a marked reduction of the witness beam density in case III, due to the impact of saturation.

The parameters for the dummy driver and the witness beam in all scenarios are summarized in Table~\ref{AWAKEparam}. Now that all the parameters are available, we can commence the simulation study of the witness beam acceleration and its associated radiation emission.
 \section{simulation results and discussion} \label{Sec4-1}
\begin{figure*}[]
	\centering
	\includegraphics[angle=0, width=.99\textwidth]{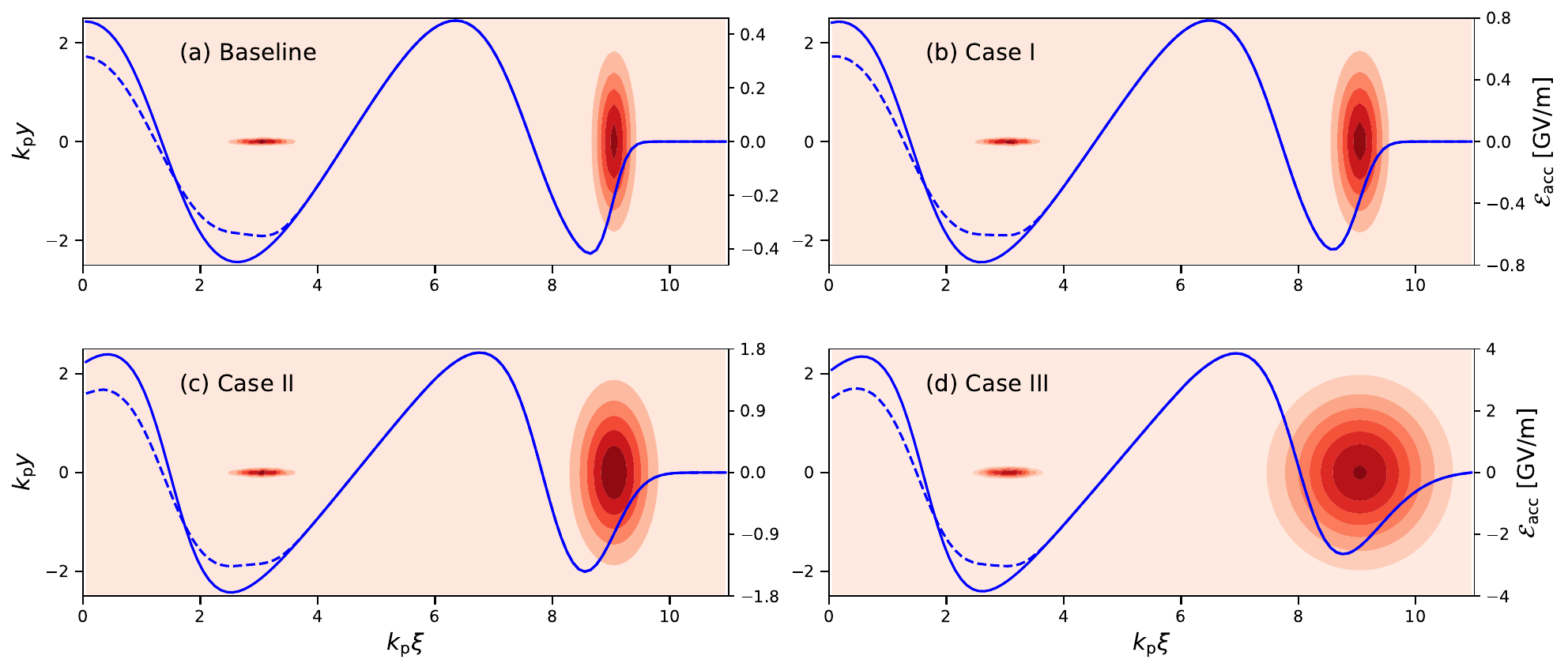}
	\caption{\footnotesize{The wakefield (blue line), the beam-loaded wakefield (dashed blue line), and contour plots of the witness beam at $k_\mathrm{p} \xi = 3$ and the dummy driver at $k_\mathrm{p} \xi= 9$ for Baseline (a), Case I (b), Case II (c) and Case III (d). Parameters are detailed in Table~\ref{AWAKEparam}. 
 }}
	\label{wakefields}
\label{singleparticle}
\end{figure*} 
PIC simulations are conducted by defining a simulation window of dimensions $(11, 5, 5) \times k_\mathrm{p}^{-1}$ in $(\mathrm{x, \, y, \, z)}$ directions. Here, $x$ denotes the longitudinal direction, and $y$ and $z$ represent the transverse directions. A spatial resolution of $(0.1, 0.05, 0.05) \times k_\mathrm{p}^{-1}$, and a time step of $5 \omega_\mathrm{p}^{-1}$ are considered in all scenarios. The number of macroparticles per cell for the plasma, driver, and witness beams are set at 4, 1, and 16, respectively. A spatial delay of $k_\mathrm{p}\xi = 6$ is introduced between the witness and driver beams to ensure that the witness beam is in the accelerating phase of the wakefield. Details of all simulation parameters are summarized in Table~\ref{AWAKEparam}. 

In Fig.~\ref{wakefields}, the wakefield (blue line), the beam-loaded wakefield (dashed blue line), and the driver and witness beams are depicted for Baseline (a), Case I (b), Case II (c), and Case III (d). It is evident that the maximum unloaded wakefields align with the $\mathcal{E}_\mathrm{peak}$ as outlined in the Table~\ref{pealwake}. On the other hand, the loaded wakefields exhibit near uniform fields, facilitating the acceleration of the witness beam in a quasi-monoenergetic manner.
\begin{figure}[]
	\centering
	\includegraphics[angle=0, width=.49\textwidth]{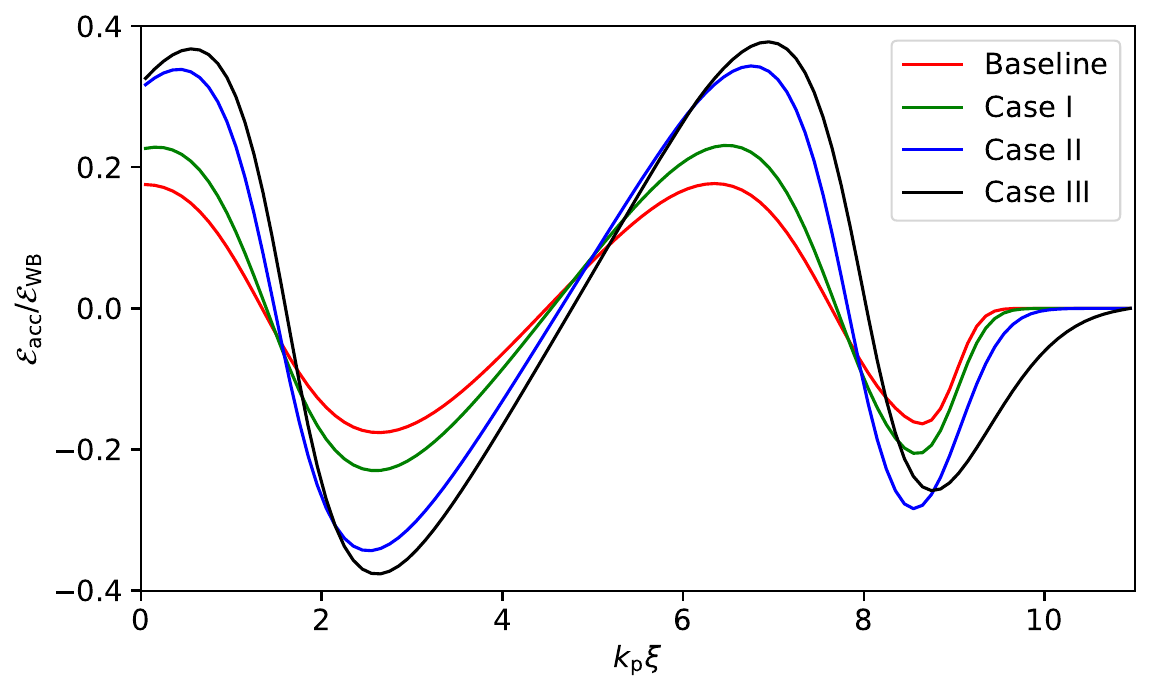}
	\caption{\footnotesize{The driver wakefields in dimensionless units.}}
	\label{comparewake}
\label{singleparticle}
\end{figure} 
A comparative plot of the driver wakefields in dimensionless units is shown in Fig.~\ref{comparewake}. The wakefields peak almost at the position of the witness beam.

\begin{figure}[]
	\centering
	\includegraphics[angle=0, width=.49\textwidth]{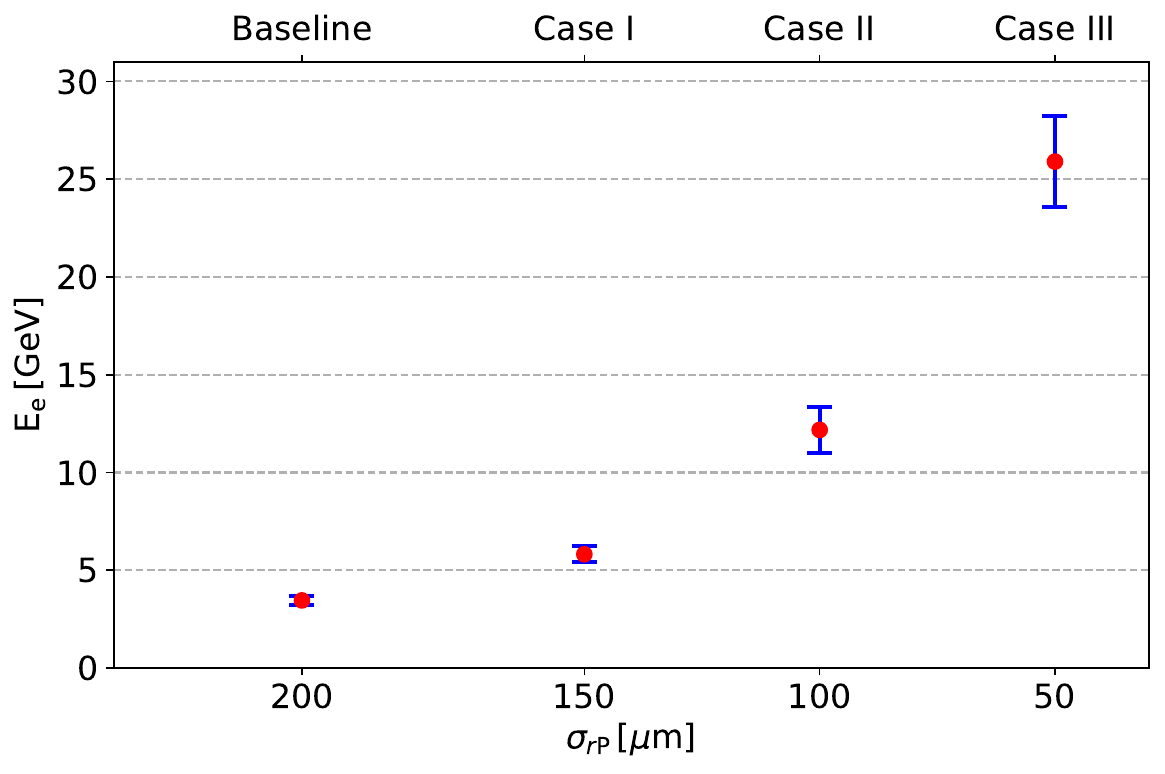}
	\caption{\footnotesize{Final energy of the witness beam (in red) along with the corresponding energy spread (in blue) for various scenarios outlined in Table~\ref{AWAKEparam}.}}
	\label{gain}
\label{EdE}
\end{figure}

In Fig.~\ref{gain} the energy gain and energy spread of the witness beam in various scenarios are depicted. As anticipated, the energy gain is significantly enhanced with increasing plasma density. The Baseline achieves 3.5 GeV with 7\% energy spread, a result comparable to those reported by Liang et al.\cite{Liang2023} The toy model simulations conducted by Olsen et. al.~\cite{Olsen2018} for a similar witness beam, albeit with charge and energy of 100 pC and 217 MeV, reveal a mean momentum of 1.67 GeV/c and an energy spread of 5.2\% over 4 m of plasma. Moreover, Fig.~\ref{gain} also illustrates that the Case I, II and III achieve final energies of 5.8 GeV, 12.2 GeV and 25.9 GeV, respectively, with energy spreads of 7\%, 9\% and 9\%, respectively. The findings indicate that the plasma density is a crucial factor for resonantly driven wakefield acceleration over a restricted length. A straightforward comparison with the Baseline reveals that augmenting the plasma density by a factor of 1.8 (as in Case I) results in a 1.7-fold increase in energy gain. Likewise, in Case II, where the plasma density increases by a factor of 4, the energy gain escalates by a factor of 3.5 compared with the Baseline. In Case III, with a plasma density increase by a factor of 16.1, the energy gain rises by a factor of 7.4. 
As previously discussed, an increase in plasma density leads to a rise in both the number of microbunches and the wakefield driven by an individual microbunch. In linear scenarios as in Baseline, Case I and Case II, this results in an accelerating field proportional to $\omega_\mathrm{p}^2$, i.e. the plasma density. However, upon reaching saturation as in Case III, the benefit of ``more microbunches'' diminishes since the wakes saturate after a certain number of microbunches. Nevertheless, the wake driven by one single microbunch still remains higher, resulting in a scaling of wakefield approximately proportional to $\omega_\mathrm{p}$, i.e. $\sqrt{n_\mathrm{p}}$.

\begin{figure}[]
	\centering
	\includegraphics[angle=0, width=.49\textwidth]{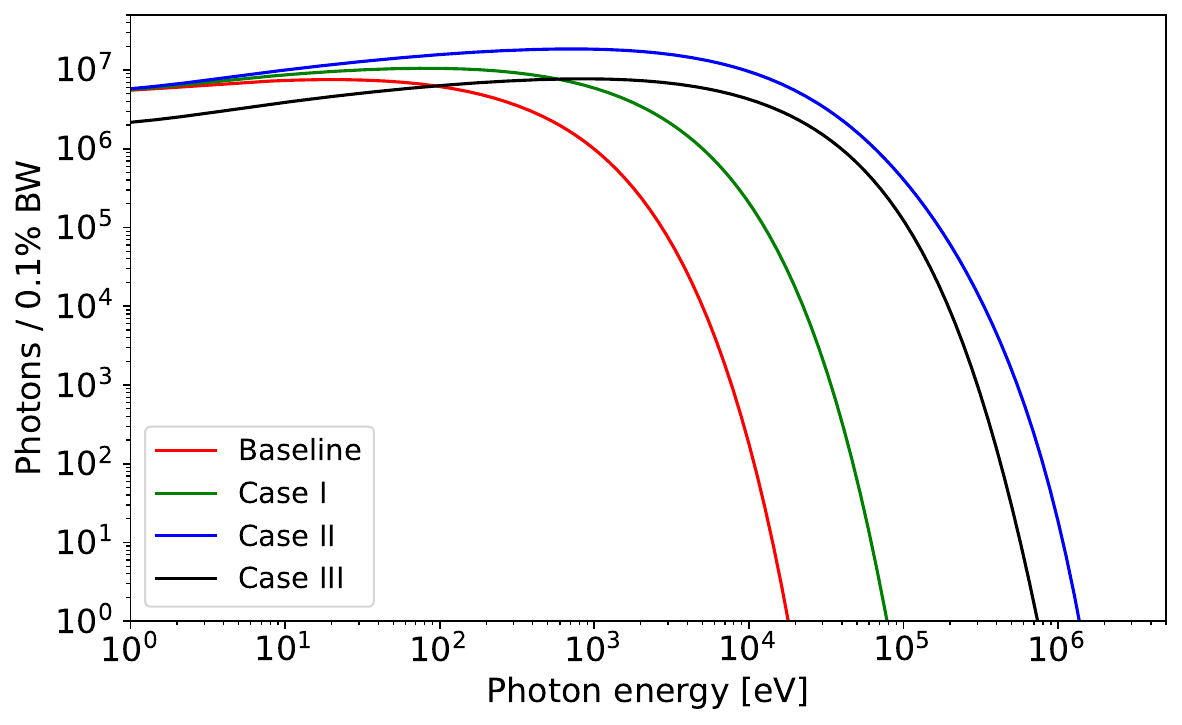}
	\caption{\footnotesize{Betatron spectra for various scenarios outlined in Table~\ref{AWAKEparam}.}}
	\label{spectrum}
\label{EdE}
\end{figure}

Figure~\ref{spectrum} displays the witness beam radiation spectra for the Baseline and three distinct cases of high plasma densities. The spectra are integrated over the 10-metre plasma length. The radiation spectrum of the Baseline is primarily in the UV to low-energy X-ray range, featuring a critical energy of $E_\mathrm{c} = 302 \, \mathrm{eV}$. The critical photon energy of the bunch is calculated as $E_\mathrm{c}  =15\sqrt{3} \left<E\right>/8$
where $\left<E\right>$ is the mean photon energy of the spectrum. At the critical energy half of the radiated energy is below it and the other half is above. The simulation results indicate the critical energies of $1.1 \, \mathrm{keV}$, $9.0 \, \mathrm{keV}$ and $9.1 \, \mathrm{keV}$ for Case I, Case II and Case III, respectively. The spectrum spans more broadly into the hard X-ray to the gamma ray range as the plasma density elevated. It is noticeable that the spectrum in Case III does not exhibit an increase in photon numbers compared to Case II. This distinct nature of Case III stems from saturation, necessitating a reduction in the witness current to maintain equivalent acceleration. This leads to a dual reduction in the witness charge—firstly due to its shorter length and secondarily because of the lower current.


The BR would exhibit a pulse length on the order of 100 femtosecond, encompassing  total photon numbers of $3.8 \times 10^9$ in Baseline, $2.2 \times 10^{10}$ in Case I, $3.7 \times 10^{11}$ in Case II and $1.5 \times 10^{11}$ in Case III.  
\section{\label{sec5}Conclusion}
This paper presents a simulation study investigating the impact of plasma density on the energy gain of the witness beam and the corresponding betatron radiation in a resonantly-driven wakefield produced by a multi-microbunch driver. The peak wakefield is computed by employing a periodic train driver in four distinct scenarios of different densities. The wakefield grows until nonlinear effects occur, which leads to a subsequent decay in the wakefield. The simulations demonstrate that the elevated plasma density significantly enhances electron energy gain. For instance, in Case III, where the plasma density is 16 times higher, the witness beam achieves about 7.4 times higher energy compared to the Baseline. Nevertheless, the nonlinear effect of wakefield decay becomes evident in Case III as the energy gain scales less than $\omega_\mathrm{p}^2$. The findings also indicate that the elevated plasma density results in the emission of more betatron photons at higher energies until wakefield saturation occurs. For example, in Case II, the plasma density is four times that of the Baseline, where the the corresponding BR extends into the hard X-ray and gamma ray range, with approximately 100 times more photons per pulse. It is noteworthy that the process of focusing the proton beam may require the implementation of some experimental techniques, particularly considering that it could inherently result in a broader divergence of the beam. This effect will be thoroughly explored in our forthcoming research endeavors.


   \begin{acknowledgments}
The authors would like to acknowledge the support from the Cockcroft Institute Core
Grant No. ST/V001612/1 and the STFC AWAKE Run 2 Grant Nos. ST/T001917/1 and
ST/X00614X/1.
\end{acknowledgments}

\section*{Data Availability Statement}
The data that support the findings of this study are available from the corresponding author upon reasonable request.
\bibliography{aipsamp}

\begin{thebibliography}{32}%
\makeatletter
\providecommand \@ifxundefined [1]{%
 \@ifx{#1\undefined}
}%
\providecommand \@ifnum [1]{%
 \ifnum #1\expandafter \@firstoftwo
 \else \expandafter \@secondoftwo
 \fi
}%
\providecommand \@ifx [1]{%
 \ifx #1\expandafter \@firstoftwo
 \else \expandafter \@secondoftwo
 \fi
}%
\providecommand \natexlab [1]{#1}%
\providecommand \enquote  [1]{``#1''}%
\providecommand \bibnamefont  [1]{#1}%
\providecommand \bibfnamefont [1]{#1}%
\providecommand \citenamefont [1]{#1}%
\providecommand \href@noop [0]{\@secondoftwo}%
\providecommand \href [0]{\begingroup \@sanitize@url \@href}%
\providecommand \@href[1]{\@@startlink{#1}\@@href}%
\providecommand \@@href[1]{\endgroup#1\@@endlink}%
\providecommand \@sanitize@url [0]{\catcode `\\12\catcode `\$12\catcode
  `\&12\catcode `\#12\catcode `\^12\catcode `\_12\catcode `\%12\relax}%
\providecommand \@@startlink[1]{}%
\providecommand \@@endlink[0]{}%
\providecommand \url  [0]{\begingroup\@sanitize@url \@url }%
\providecommand \@url [1]{\endgroup\@href {#1}{\urlprefix }}%
\providecommand \urlprefix  [0]{URL }%
\providecommand \Eprint [0]{\href }%
\providecommand \doibase [0]{http://dx.doi.org/}%
\providecommand \selectlanguage [0]{\@gobble}%
\providecommand \bibinfo  [0]{\@secondoftwo}%
\providecommand \bibfield  [0]{\@secondoftwo}%
\providecommand \translation [1]{[#1]}%
\providecommand \BibitemOpen [0]{}%
\providecommand \bibitemStop [0]{}%
\providecommand \bibitemNoStop [0]{.\EOS\space}%
\providecommand \EOS [0]{\spacefactor3000\relax}%
\providecommand \BibitemShut  [1]{\csname bibitem#1\endcsname}%
\let\auto@bib@innerbib\@empty
\bibitem [{\citenamefont {Esarey}, \citenamefont {Schroeder},\ and\
  \citenamefont {Leemans}(2009)}]{Esarey2009}%
  \BibitemOpen
  \bibfield  {author} {\bibinfo {author} {\bibfnamefont {E.}~\bibnamefont
  {Esarey}}, \bibinfo {author} {\bibfnamefont {C.~B.}\ \bibnamefont
  {Schroeder}}, \ and\ \bibinfo {author} {\bibfnamefont {W.~P.}\ \bibnamefont
  {Leemans}},\ }\bibfield  {title} {\enquote {\bibinfo {title} {Physics of
  laser-driven plasma-based electron accelerators},}\ }\href {\doibase
  10.1103/RevModPhys.81.1229} {\bibfield  {journal} {\bibinfo  {journal} {Rev.
  Mod. Phys.}\ }\textbf {\bibinfo {volume} {81}} (\bibinfo {year} {2009}),\
  10.1103/RevModPhys.81.1229}\BibitemShut {NoStop}%
\bibitem [{\citenamefont {Hogan}(2016)}]{Hogan2016}%
  \BibitemOpen
  \bibfield  {author} {\bibinfo {author} {\bibfnamefont {M.~J.}\ \bibnamefont
  {Hogan}},\ }\bibfield  {title} {\enquote {\bibinfo {title} {Electron and
  positron beam–driven plasma acceleration},}\ }\href {\doibase
  10.1142/S1793626816300036} {\bibfield  {journal} {\bibinfo  {journal}
  {Reviews of Accelerator Science and Technology}\ }\textbf {\bibinfo {volume}
  {9}},\ \bibinfo {pages} {63--83} (\bibinfo {year} {2016})}\BibitemShut
  {NoStop}%
\bibitem [{\citenamefont {Adli}\ and\ \citenamefont {Muggli}(2016)}]{Adli2016}%
  \BibitemOpen
  \bibfield  {author} {\bibinfo {author} {\bibfnamefont {E.}~\bibnamefont
  {Adli}}\ and\ \bibinfo {author} {\bibfnamefont {P.}~\bibnamefont {Muggli}},\
  }\bibfield  {title} {\enquote {\bibinfo {title} {Proton-beam-driven plasma
  acceleration},}\ }\href {\doibase 10.1142/S1793626816300048} {\bibfield
  {journal} {\bibinfo  {journal} {Reviews of Accelerator Science and
  Technology}\ }\textbf {\bibinfo {volume} {9}},\ \bibinfo {pages} {85--104}
  (\bibinfo {year} {2016})}\BibitemShut {NoStop}%
\bibitem [{\citenamefont {Esarey}\ \emph {et~al.}(2002)\citenamefont {Esarey},
  \citenamefont {Shadwick}, \citenamefont {Catravas},\ and\ \citenamefont
  {Leemans}}]{Esarey2002}%
  \BibitemOpen
  \bibfield  {author} {\bibinfo {author} {\bibfnamefont {E.}~\bibnamefont
  {Esarey}}, \bibinfo {author} {\bibfnamefont {B.~A.}\ \bibnamefont
  {Shadwick}}, \bibinfo {author} {\bibfnamefont {P.}~\bibnamefont {Catravas}},
  \ and\ \bibinfo {author} {\bibfnamefont {W.~P.}\ \bibnamefont {Leemans}},\
  }\bibfield  {title} {\enquote {\bibinfo {title} {Synchrotron radiation from
  electron beams in plasma-focusing channels},}\ }\href {\doibase
  10.1103/PhysRevE.65.056505} {\bibfield  {journal} {\bibinfo  {journal} {Phys.
  Rev. E}\ }\textbf {\bibinfo {volume} {65}},\ \bibinfo {pages} {056505}
  (\bibinfo {year} {2002})}\BibitemShut {NoStop}%
\bibitem [{\citenamefont {Kostyukov}, \citenamefont {Kiselev},\ and\
  \citenamefont {Pukhov}(2003)}]{Kostyukov2003}%
  \BibitemOpen
  \bibfield  {author} {\bibinfo {author} {\bibfnamefont {I.}~\bibnamefont
  {Kostyukov}}, \bibinfo {author} {\bibfnamefont {S.}~\bibnamefont {Kiselev}},
  \ and\ \bibinfo {author} {\bibfnamefont {A.}~\bibnamefont {Pukhov}},\
  }\bibfield  {title} {\enquote {\bibinfo {title} {X-ray generation in an ion
  channel},}\ }\href {\doibase 10.1063/1.1624605} {\bibfield  {journal}
  {\bibinfo  {journal} {Phys. Plasmas}\ }\textbf {\bibinfo {volume} {10}},\
  \bibinfo {pages} {4818--4828} (\bibinfo {year} {2003})}\BibitemShut {NoStop}%
\bibitem [{\citenamefont {Wang}\ \emph {et~al.}(2002)\citenamefont {Wang},
  \citenamefont {Clayton}, \citenamefont {Blue}, \citenamefont {Dodd},
  \citenamefont {Marsh}, \citenamefont {Mori}, \citenamefont {Joshi},
  \citenamefont {Lee}, \citenamefont {Muggli}, \citenamefont {Katsouleas},
  \citenamefont {Decker}, \citenamefont {Hogan}, \citenamefont {Iverson},
  \citenamefont {Raimondi}, \citenamefont {Walz}, \citenamefont {Siemann},\
  and\ \citenamefont {Assmann}}]{PhysRevLett.88.135004}%
  \BibitemOpen
  \bibfield  {author} {\bibinfo {author} {\bibfnamefont {S.}~\bibnamefont
  {Wang}}, \bibinfo {author} {\bibfnamefont {C.~E.}\ \bibnamefont {Clayton}},
  \bibinfo {author} {\bibfnamefont {B.~E.}\ \bibnamefont {Blue}}, \bibinfo
  {author} {\bibfnamefont {E.~S.}\ \bibnamefont {Dodd}}, \bibinfo {author}
  {\bibfnamefont {K.~A.}\ \bibnamefont {Marsh}}, \bibinfo {author}
  {\bibfnamefont {W.~B.}\ \bibnamefont {Mori}}, \bibinfo {author}
  {\bibfnamefont {C.}~\bibnamefont {Joshi}}, \bibinfo {author} {\bibfnamefont
  {S.}~\bibnamefont {Lee}}, \bibinfo {author} {\bibfnamefont {P.}~\bibnamefont
  {Muggli}}, \bibinfo {author} {\bibfnamefont {T.}~\bibnamefont {Katsouleas}},
  \bibinfo {author} {\bibfnamefont {F.~J.}\ \bibnamefont {Decker}}, \bibinfo
  {author} {\bibfnamefont {M.~J.}\ \bibnamefont {Hogan}}, \bibinfo {author}
  {\bibfnamefont {R.~H.}\ \bibnamefont {Iverson}}, \bibinfo {author}
  {\bibfnamefont {P.}~\bibnamefont {Raimondi}}, \bibinfo {author}
  {\bibfnamefont {D.}~\bibnamefont {Walz}}, \bibinfo {author} {\bibfnamefont
  {R.}~\bibnamefont {Siemann}}, \ and\ \bibinfo {author} {\bibfnamefont
  {R.}~\bibnamefont {Assmann}},\ }\bibfield  {title} {\enquote {\bibinfo
  {title} {X-ray emission from betatron motion in a plasma wiggler},}\ }\href
  {\doibase 10.1103/PhysRevLett.88.135004} {\bibfield  {journal} {\bibinfo
  {journal} {Phys. Rev. Lett.}\ }\textbf {\bibinfo {volume} {88}},\ \bibinfo
  {pages} {135004} (\bibinfo {year} {2002})}\BibitemShut {NoStop}%
\bibitem [{\citenamefont {Rousse}\ \emph {et~al.}(2004)\citenamefont {Rousse},
  \citenamefont {Phuoc}, \citenamefont {Shah}, \citenamefont {Pukhov},
  \citenamefont {Lefebvre}, \citenamefont {Malka}, \citenamefont {Kiselev},
  \citenamefont {Burgy}, \citenamefont {Rousseau}, \citenamefont {Umstadter},\
  and\ \citenamefont {Hulin}}]{Rousse2004}%
  \BibitemOpen
  \bibfield  {author} {\bibinfo {author} {\bibfnamefont {A.}~\bibnamefont
  {Rousse}}, \bibinfo {author} {\bibfnamefont {K.~T.}\ \bibnamefont {Phuoc}},
  \bibinfo {author} {\bibfnamefont {R.}~\bibnamefont {Shah}}, \bibinfo {author}
  {\bibfnamefont {A.}~\bibnamefont {Pukhov}}, \bibinfo {author} {\bibfnamefont
  {E.}~\bibnamefont {Lefebvre}}, \bibinfo {author} {\bibfnamefont
  {V.}~\bibnamefont {Malka}}, \bibinfo {author} {\bibfnamefont
  {S.}~\bibnamefont {Kiselev}}, \bibinfo {author} {\bibfnamefont
  {F.}~\bibnamefont {Burgy}}, \bibinfo {author} {\bibfnamefont {J.-P.}\
  \bibnamefont {Rousseau}}, \bibinfo {author} {\bibfnamefont {D.}~\bibnamefont
  {Umstadter}}, \ and\ \bibinfo {author} {\bibfnamefont {D.}~\bibnamefont
  {Hulin}},\ }\bibfield  {title} {\enquote {\bibinfo {title} {Production of a
  kev x-ray beam from synchrotron radiation in relativistic laser-plasma
  interaction},}\ }\href {\doibase 10.1103/PhysRevLett.93.135005} {\bibfield
  {journal} {\bibinfo  {journal} {Phys. Rev. Lett.}\ }\textbf {\bibinfo
  {volume} {93}},\ \bibinfo {pages} {135005} (\bibinfo {year}
  {2004})}\BibitemShut {NoStop}%
\bibitem [{\citenamefont {Kneip}\ \emph {et~al.}(2010)\citenamefont {Kneip},
  \citenamefont {McGuffey}, \citenamefont {Martins}, \citenamefont {Martins},
  \citenamefont {Bellei}, \citenamefont {Chvykov}, \citenamefont {Fonseca},
  \citenamefont {Huntington}, \citenamefont {Kalintchenko},\ and\ \citenamefont
  {et. al.}}]{Kneip2010}%
  \BibitemOpen
  \bibfield  {author} {\bibinfo {author} {\bibfnamefont {S.}~\bibnamefont
  {Kneip}}, \bibinfo {author} {\bibfnamefont {C.}~\bibnamefont {McGuffey}},
  \bibinfo {author} {\bibfnamefont {J.~L.}\ \bibnamefont {Martins}}, \bibinfo
  {author} {\bibfnamefont {S.~F.}\ \bibnamefont {Martins}}, \bibinfo {author}
  {\bibfnamefont {C.}~\bibnamefont {Bellei}}, \bibinfo {author} {\bibfnamefont
  {V.}~\bibnamefont {Chvykov}}, \bibinfo {author} {\bibfnamefont {F.~D.~R.}\
  \bibnamefont {Fonseca}}, \bibinfo {author} {\bibfnamefont {C.}~\bibnamefont
  {Huntington}}, \bibinfo {author} {\bibfnamefont {G.}~\bibnamefont
  {Kalintchenko}}, \ and\ \bibinfo {author} {\bibnamefont {et. al.}},\
  }\bibfield  {title} {\enquote {\bibinfo {title} {Bright spatially coherent
  synchrotron x-rays from a table-top source},}\ }\href {\doibase
  10.1038/nphys1789} {\bibfield  {journal} {\bibinfo  {journal} {Nature
  Physics}\ }\textbf {\bibinfo {volume} {6}},\ \bibinfo {pages} {980--983}
  (\bibinfo {year} {2010})}\BibitemShut {NoStop}%
\bibitem [{\citenamefont {Cipiccia}\ \emph {et~al.}(2011)\citenamefont
  {Cipiccia}, \citenamefont {Islam}, \citenamefont {Ersfeld}, \citenamefont
  {Shanks}, \citenamefont {Brunetti}, \citenamefont {Vieux}, \citenamefont
  {Yang}, \citenamefont {Issac}, \citenamefont {M.Wiggins}, \citenamefont
  {H.Welsh},\ and\ \citenamefont {et~al}}]{Cipiccia2011}%
  \BibitemOpen
  \bibfield  {author} {\bibinfo {author} {\bibfnamefont {S.}~\bibnamefont
  {Cipiccia}}, \bibinfo {author} {\bibfnamefont {M.~R.}\ \bibnamefont {Islam}},
  \bibinfo {author} {\bibfnamefont {B.}~\bibnamefont {Ersfeld}}, \bibinfo
  {author} {\bibfnamefont {R.~P.}\ \bibnamefont {Shanks}}, \bibinfo {author}
  {\bibfnamefont {E.}~\bibnamefont {Brunetti}}, \bibinfo {author}
  {\bibfnamefont {G.}~\bibnamefont {Vieux}}, \bibinfo {author} {\bibfnamefont
  {X.}~\bibnamefont {Yang}}, \bibinfo {author} {\bibfnamefont {R.~C.}\
  \bibnamefont {Issac}}, \bibinfo {author} {\bibfnamefont {S.}~\bibnamefont
  {M.Wiggins}}, \bibinfo {author} {\bibfnamefont {G.}~\bibnamefont {H.Welsh}},
  \ and\ \bibinfo {author} {\bibnamefont {et~al}},\ }\bibfield  {title}
  {\enquote {\bibinfo {title} {Single shot phase contrast imaging using
  laser-produced betatron x-ray beams},}\ }\href {\doibase 10.1038/nphys2090}
  {\bibfield  {journal} {\bibinfo  {journal} {Nature Phys}\ }\textbf {\bibinfo
  {volume} {36}},\ \bibinfo {pages} {867–871} (\bibinfo {year}
  {2011})}\BibitemShut {NoStop}%
\bibitem [{\citenamefont {Fourmaux}\ \emph {et~al.}(2011)\citenamefont
  {Fourmaux}, \citenamefont {Corde}, \citenamefont {Phuoc}, \citenamefont
  {Lassonde}, \citenamefont {Lebrun}, \citenamefont {Payeur}, \citenamefont
  {Martin}, \citenamefont {Sebban}, \citenamefont {Malka}, \citenamefont
  {Rousse},\ and\ \citenamefont {Kieffer}}]{Fourmaux2011}%
  \BibitemOpen
  \bibfield  {author} {\bibinfo {author} {\bibfnamefont {S.}~\bibnamefont
  {Fourmaux}}, \bibinfo {author} {\bibfnamefont {S.}~\bibnamefont {Corde}},
  \bibinfo {author} {\bibfnamefont {K.~T.}\ \bibnamefont {Phuoc}}, \bibinfo
  {author} {\bibfnamefont {P.}~\bibnamefont {Lassonde}}, \bibinfo {author}
  {\bibfnamefont {G.}~\bibnamefont {Lebrun}}, \bibinfo {author} {\bibfnamefont
  {S.}~\bibnamefont {Payeur}}, \bibinfo {author} {\bibfnamefont
  {F.}~\bibnamefont {Martin}}, \bibinfo {author} {\bibfnamefont
  {S.}~\bibnamefont {Sebban}}, \bibinfo {author} {\bibfnamefont
  {V.}~\bibnamefont {Malka}}, \bibinfo {author} {\bibfnamefont
  {A.}~\bibnamefont {Rousse}}, \ and\ \bibinfo {author} {\bibfnamefont {J.~C.}\
  \bibnamefont {Kieffer}},\ }\bibfield  {title} {\enquote {\bibinfo {title}
  {Single shot phase contrast imaging using laser-produced betatron x-ray
  beams},}\ }\href {\doibase 10.1364/OL.36.002426} {\bibfield  {journal}
  {\bibinfo  {journal} {Opt Lett.}\ }\textbf {\bibinfo {volume} {36}},\
  \bibinfo {pages} {2426--8} (\bibinfo {year} {2011})}\BibitemShut {NoStop}%
\bibitem [{\citenamefont {Kneip}\ \emph {et~al.}(2012)\citenamefont {Kneip},
  \citenamefont {McGuffey}, \citenamefont {Martins}, \citenamefont {Bloom},
  \citenamefont {Chvykov}, \citenamefont {Dollar}, \citenamefont {Fonseca},
  \citenamefont {Jolly}, \citenamefont {Kalintchenko}, \citenamefont
  {Krushelnick},\ and\ \citenamefont {et~al}}]{Kneip2012}%
  \BibitemOpen
  \bibfield  {author} {\bibinfo {author} {\bibfnamefont {S.}~\bibnamefont
  {Kneip}}, \bibinfo {author} {\bibfnamefont {C.}~\bibnamefont {McGuffey}},
  \bibinfo {author} {\bibfnamefont {J.~L.}\ \bibnamefont {Martins}}, \bibinfo
  {author} {\bibfnamefont {M.~S.}\ \bibnamefont {Bloom}}, \bibinfo {author}
  {\bibfnamefont {V.}~\bibnamefont {Chvykov}}, \bibinfo {author} {\bibfnamefont
  {F.}~\bibnamefont {Dollar}}, \bibinfo {author} {\bibfnamefont
  {R.}~\bibnamefont {Fonseca}}, \bibinfo {author} {\bibfnamefont
  {S.}~\bibnamefont {Jolly}}, \bibinfo {author} {\bibfnamefont
  {G.}~\bibnamefont {Kalintchenko}}, \bibinfo {author} {\bibfnamefont
  {K.}~\bibnamefont {Krushelnick}}, \ and\ \bibinfo {author} {\bibnamefont
  {et~al}},\ }\bibfield  {title} {\enquote {\bibinfo {title} {Characterization
  of transverse beam emittance of electrons from a laser-plasma wakefield
  accelerator in the bubble regime using betatron x-ray radiation},}\ }\href
  {\doibase 10.1103/PhysRevSTAB.15.021302} {\bibfield  {journal} {\bibinfo
  {journal} {Physical Review Accelerators and Beams}\ }\textbf {\bibinfo
  {volume} {15}},\ \bibinfo {pages} {021302} (\bibinfo {year}
  {2012})}\BibitemShut {NoStop}%
\bibitem [{\citenamefont {A.Köhler}\ \emph {et~al.}(2016)\citenamefont
  {A.Köhler}, \citenamefont {J.P.Couperus}, \citenamefont {O.Zarini},
  \citenamefont {A.Jochmann}, \citenamefont {A.Irman},\ and\ \citenamefont
  {U.Schramm}}]{Kohler2016}%
  \BibitemOpen
  \bibfield  {author} {\bibinfo {author} {\bibnamefont {A.Köhler}}, \bibinfo
  {author} {\bibnamefont {J.P.Couperus}}, \bibinfo {author} {\bibnamefont
  {O.Zarini}}, \bibinfo {author} {\bibnamefont {A.Jochmann}}, \bibinfo {author}
  {\bibnamefont {A.Irman}}, \ and\ \bibinfo {author} {\bibnamefont
  {U.Schramm}},\ }\bibfield  {title} {\enquote {\bibinfo {title} {Single-shot
  betatron source size measurement from a laser-wakefield accelerator},}\
  }\href {\doibase 10.1016/j.nima.2016.02.031} {\bibfield  {journal} {\bibinfo
  {journal} {Nuclear Instruments and Methods in Physics Research A}\ }\textbf
  {\bibinfo {volume} {829}},\ \bibinfo {pages} {265–269} (\bibinfo {year}
  {2016})}\BibitemShut {NoStop}%
\bibitem [{\citenamefont {Curcio}\ \emph {et~al.}(2017)\citenamefont {Curcio},
  \citenamefont {Anania}, \citenamefont {Bisesto}, \citenamefont {Chiadroni},
  \citenamefont {Cianchi}, \citenamefont {Ferrario}, \citenamefont {Filippi},
  \citenamefont {Giulietti}, \citenamefont {Marocchino}, \citenamefont
  {Petrarca},\ and\ \citenamefont {et~al}}]{Curcio2017}%
  \BibitemOpen
  \bibfield  {author} {\bibinfo {author} {\bibfnamefont {A.}~\bibnamefont
  {Curcio}}, \bibinfo {author} {\bibfnamefont {M.}~\bibnamefont {Anania}},
  \bibinfo {author} {\bibfnamefont {F.}~\bibnamefont {Bisesto}}, \bibinfo
  {author} {\bibfnamefont {E.}~\bibnamefont {Chiadroni}}, \bibinfo {author}
  {\bibfnamefont {A.}~\bibnamefont {Cianchi}}, \bibinfo {author} {\bibfnamefont
  {M.}~\bibnamefont {Ferrario}}, \bibinfo {author} {\bibfnamefont
  {F.}~\bibnamefont {Filippi}}, \bibinfo {author} {\bibfnamefont
  {D.}~\bibnamefont {Giulietti}}, \bibinfo {author} {\bibfnamefont
  {A.}~\bibnamefont {Marocchino}}, \bibinfo {author} {\bibfnamefont
  {M.}~\bibnamefont {Petrarca}}, \ and\ \bibinfo {author} {\bibnamefont
  {et~al}},\ }\bibfield  {title} {\enquote {\bibinfo {title} {Trace-space
  reconstruction of low-emittance electron beams through betatron radiation in
  laser-plasma accelerators},}\ }\href {\doibase
  10.1103/PhysRevAccelBeams.20.012801} {\bibfield  {journal} {\bibinfo
  {journal} {Physical Review Accelerators and Beams}\ }\textbf {\bibinfo
  {volume} {20}},\ \bibinfo {pages} {012801} (\bibinfo {year}
  {2017})}\BibitemShut {NoStop}%
\bibitem [{\citenamefont {Shpakov}\ \emph {et~al.}(2016)\citenamefont
  {Shpakov}, \citenamefont {M.P.Anania}, \citenamefont {A.Biagioni},
  \citenamefont {E.Chiadroni}, \citenamefont {A.Cianchi}, \citenamefont
  {A.Curcio}, \citenamefont {Dabagov}, \citenamefont {M.Ferrario},
  \citenamefont {F.Filippi}, \citenamefont {A.Marocchino},\ and\ \citenamefont
  {et~al}}]{Shpakov2016}%
  \BibitemOpen
  \bibfield  {author} {\bibinfo {author} {\bibfnamefont {V.}~\bibnamefont
  {Shpakov}}, \bibinfo {author} {\bibnamefont {M.P.Anania}}, \bibinfo {author}
  {\bibnamefont {A.Biagioni}}, \bibinfo {author} {\bibnamefont {E.Chiadroni}},
  \bibinfo {author} {\bibnamefont {A.Cianchi}}, \bibinfo {author} {\bibnamefont
  {A.Curcio}}, \bibinfo {author} {\bibfnamefont {S.}~\bibnamefont {Dabagov}},
  \bibinfo {author} {\bibnamefont {M.Ferrario}}, \bibinfo {author}
  {\bibnamefont {F.Filippi}}, \bibinfo {author} {\bibnamefont {A.Marocchino}},
  \ and\ \bibinfo {author} {\bibnamefont {et~al}},\ }\bibfield  {title}
  {\enquote {\bibinfo {title} {Betatron radiation based diagnostics for plasma
  wakefield accelerated electron beams at the sparclab test facility},}\ }\href
  {\doibase 10.1016/j.nima.2016.02.074} {\bibfield  {journal} {\bibinfo
  {journal} {Nuclear Instruments and Methods in Physics Research A}\ }\textbf
  {\bibinfo {volume} {829}},\ \bibinfo {pages} {330--333} (\bibinfo {year}
  {2016})}\BibitemShut {NoStop}%
\bibitem [{\citenamefont {Claveria}\ \emph {et~al.}(2019)\citenamefont
  {Claveria}, \citenamefont {Adli}, \citenamefont {Amorim}, \citenamefont {An},
  \citenamefont {Clayton}, \citenamefont {Corde}, \citenamefont {Gessner},
  \citenamefont {Hogan}, \citenamefont {Joshi}, \citenamefont {Kononenko},\
  and\ \citenamefont {et~al}}]{Claveria2019}%
  \BibitemOpen
  \bibfield  {author} {\bibinfo {author} {\bibfnamefont {P.~S.~M.}\
  \bibnamefont {Claveria}}, \bibinfo {author} {\bibfnamefont {E.}~\bibnamefont
  {Adli}}, \bibinfo {author} {\bibfnamefont {L.~D.}\ \bibnamefont {Amorim}},
  \bibinfo {author} {\bibfnamefont {W.}~\bibnamefont {An}}, \bibinfo {author}
  {\bibfnamefont {C.~E.}\ \bibnamefont {Clayton}}, \bibinfo {author}
  {\bibfnamefont {S.}~\bibnamefont {Corde}}, \bibinfo {author} {\bibfnamefont
  {S.}~\bibnamefont {Gessner}}, \bibinfo {author} {\bibfnamefont {M.~J.}\
  \bibnamefont {Hogan}}, \bibinfo {author} {\bibfnamefont {C.}~\bibnamefont
  {Joshi}}, \bibinfo {author} {\bibfnamefont {O.}~\bibnamefont {Kononenko}}, \
  and\ \bibinfo {author} {\bibnamefont {et~al}},\ }\bibfield  {title} {\enquote
  {\bibinfo {title} {Betatron radiation and emittance growth in plasma
  wakefield accelerators},}\ }\href {\doibase 10.1098/rsta.2018.0173}
  {\bibfield  {journal} {\bibinfo  {journal} {Philosophical Transactions of the
  Royal Society A}\ }\textbf {\bibinfo {volume} {377}},\ \bibinfo {pages}
  {20180173} (\bibinfo {year} {2019})}\BibitemShut {NoStop}%
\bibitem [{\citenamefont {Williamson}\ \emph {et~al.}(2020)\citenamefont
  {Williamson}, \citenamefont {Xia}, \citenamefont {Gessner}, \citenamefont
  {Petrenko}, \citenamefont {Farmer},\ and\ \citenamefont
  {Pukhov}}]{Williamson2020}%
  \BibitemOpen
  \bibfield  {author} {\bibinfo {author} {\bibfnamefont {B.}~\bibnamefont
  {Williamson}}, \bibinfo {author} {\bibfnamefont {G.}~\bibnamefont {Xia}},
  \bibinfo {author} {\bibfnamefont {S.}~\bibnamefont {Gessner}}, \bibinfo
  {author} {\bibfnamefont {A.}~\bibnamefont {Petrenko}}, \bibinfo {author}
  {\bibfnamefont {J.}~\bibnamefont {Farmer}}, \ and\ \bibinfo {author}
  {\bibfnamefont {A.}~\bibnamefont {Pukhov}},\ }\bibfield  {title} {\enquote
  {\bibinfo {title} {Betatron radiation diagnostics for {AWAKE} run 2},}\
  }\href {\doibase 10.1016/j.nima.2020.164076} {\bibfield  {journal} {\bibinfo
  {journal} {Nucl. Instrum. Methods Phys. Res. A}\ }\textbf {\bibinfo {volume}
  {971}},\ \bibinfo {pages} {164076} (\bibinfo {year} {2020})}\BibitemShut
  {NoStop}%
\bibitem [{\citenamefont {Liang}\ \emph {et~al.}(2023)\citenamefont {Liang},
  \citenamefont {Saberi}, \citenamefont {Xia}, \citenamefont {Farmer},\ and\
  \citenamefont {Pukhov}}]{Liang2023}%
  \BibitemOpen
  \bibfield  {author} {\bibinfo {author} {\bibfnamefont {L.}~\bibnamefont
  {Liang}}, \bibinfo {author} {\bibfnamefont {H.}~\bibnamefont {Saberi}},
  \bibinfo {author} {\bibfnamefont {G.}~\bibnamefont {Xia}}, \bibinfo {author}
  {\bibfnamefont {J.~P.}\ \bibnamefont {Farmer}}, \ and\ \bibinfo {author}
  {\bibfnamefont {A.}~\bibnamefont {Pukhov}},\ }\bibfield  {title} {\enquote
  {\bibinfo {title} {Characteristics of betatron radiation in {AWAKE} run 2
  experiment},}\ }\href {\doibase 10.1017/S0022377823000491} {\bibfield
  {journal} {\bibinfo  {journal} {Journal of Plasma Physics}\ }\textbf
  {\bibinfo {volume} {89}},\ \bibinfo {pages} {965890301} (\bibinfo {year}
  {2023})}\BibitemShut {NoStop}%
\bibitem [{\citenamefont {Caldwell}\ and\ \citenamefont
  {Lotov}(2011)}]{Caldwell2011}%
  \BibitemOpen
  \bibfield  {author} {\bibinfo {author} {\bibfnamefont {A.}~\bibnamefont
  {Caldwell}}\ and\ \bibinfo {author} {\bibfnamefont {K.}~\bibnamefont
  {Lotov}},\ }\bibfield  {title} {\enquote {\bibinfo {title} {Plasma wakefield
  acceleration with a modulated proton bunch},}\ }\href {\doibase
  10.1063/1.3641973} {\bibfield  {journal} {\bibinfo  {journal} {Physics of
  Plasmas}\ }\textbf {\bibinfo {volume} {201819}},\ \bibinfo {pages} {103101}
  (\bibinfo {year} {2011})}\BibitemShut {NoStop}%
\bibitem [{\citenamefont {Adli}\ and\ \citenamefont {et.
  al.}(2018)}]{Adli2018}%
  \BibitemOpen
  \bibfield  {author} {\bibinfo {author} {\bibfnamefont {E.}~\bibnamefont
  {Adli}}\ and\ \bibinfo {author} {\bibnamefont {et. al.}} (\bibinfo
  {collaboration} {AWAKE Collaboration}),\ }\bibfield  {title} {\enquote
  {\bibinfo {title} {Acceleration of electrons in the plasma wakefield of a
  proton bunch},}\ }\href {\doibase 10.1038/s41586-018-0485-4} {\bibfield
  {journal} {\bibinfo  {journal} {Nature}\ }\textbf {\bibinfo {volume} {561}},\
  \bibinfo {pages} {363–367} (\bibinfo {year} {2018})}\BibitemShut {NoStop}%
\bibitem [{\citenamefont {Gschwendtner}\ and\ \citenamefont {et.
  al.}(2022)}]{Gschwendtner2022}%
  \BibitemOpen
  \bibfield  {author} {\bibinfo {author} {\bibfnamefont {E.}~\bibnamefont
  {Gschwendtner}}\ and\ \bibinfo {author} {\bibnamefont {et. al.}} (\bibinfo
  {collaboration} {AWAKE Collaboration}),\ }\bibfield  {title} {\enquote
  {\bibinfo {title} {The {AWAKE} run 2 programme and beyond},}\ }\href
  {\doibase 10.3390/sym14081680} {\bibfield  {journal} {\bibinfo  {journal}
  {Symmetry}\ }\textbf {\bibinfo {volume} {14}},\ \bibinfo {pages} {1680}
  (\bibinfo {year} {2022})}\BibitemShut {NoStop}%
\bibitem [{\citenamefont {Chappell}, \citenamefont {Caldwell},\ and\
  \citenamefont {Wing}(2019)}]{Chappell2019}%
  \BibitemOpen
  \bibfield  {author} {\bibinfo {author} {\bibfnamefont {J.}~\bibnamefont
  {Chappell}}, \bibinfo {author} {\bibfnamefont {A.~C.}\ \bibnamefont
  {Caldwell}}, \ and\ \bibinfo {author} {\bibfnamefont {M.}~\bibnamefont
  {Wing}},\ }\bibfield  {title} {\enquote {\bibinfo {title} {{A compact
  electron injector for the EIC based on plasma wakefields driven by the
  RHIC-EIC proton beam}},}\ }\href {\doibase 10.22323/1.352.0219} {\bibfield
  {journal} {\bibinfo  {journal} {PoS}\ }\textbf {\bibinfo {volume}
  {DIS2019}},\ \bibinfo {pages} {219} (\bibinfo {year} {2019})}\BibitemShut
  {NoStop}%
\bibitem [{\citenamefont {Caldwell}\ \emph {et~al.}(2009)\citenamefont
  {Caldwell}, \citenamefont {Lotov}, \citenamefont {Pukhov},\ and\
  \citenamefont {Simon}}]{Caldwell2009}%
  \BibitemOpen
  \bibfield  {author} {\bibinfo {author} {\bibfnamefont {A.}~\bibnamefont
  {Caldwell}}, \bibinfo {author} {\bibfnamefont {K.}~\bibnamefont {Lotov}},
  \bibinfo {author} {\bibfnamefont {A.}~\bibnamefont {Pukhov}}, \ and\ \bibinfo
  {author} {\bibfnamefont {F.}~\bibnamefont {Simon}},\ }\bibfield  {title}
  {\enquote {\bibinfo {title} {Proton-driven plasma-wakefield acceleration},}\
  }\href {https://www.nature.com/articles/nphys1248} {\bibfield  {journal}
  {\bibinfo  {journal} {Nat. Phys.}\ }\textbf {\bibinfo {volume} {5}},\
  \bibinfo {pages} {363–367} (\bibinfo {year} {2009})}\BibitemShut {NoStop}%
\bibitem [{\citenamefont {Xia}\ \emph {et~al.}(2014)\citenamefont {Xia},
  \citenamefont {Mete}, \citenamefont {Aimidula}, \citenamefont {Welsch},
  \citenamefont {Chattopadhyay}, \citenamefont {Mandry},\ and\ \citenamefont
  {M.Wing}}]{Xia2014}%
  \BibitemOpen
  \bibfield  {author} {\bibinfo {author} {\bibfnamefont {G.}~\bibnamefont
  {Xia}}, \bibinfo {author} {\bibfnamefont {O.}~\bibnamefont {Mete}}, \bibinfo
  {author} {\bibfnamefont {A.}~\bibnamefont {Aimidula}}, \bibinfo {author}
  {\bibfnamefont {C.~P.}\ \bibnamefont {Welsch}}, \bibinfo {author}
  {\bibfnamefont {S.}~\bibnamefont {Chattopadhyay}}, \bibinfo {author}
  {\bibfnamefont {S.}~\bibnamefont {Mandry}}, \ and\ \bibinfo {author}
  {\bibnamefont {M.Wing}},\ }\bibfield  {title} {\enquote {\bibinfo {title}
  {Collider design issues based on proton-driven plasma wakefield
  acceleration},}\ }\href
  {https://www.sciencedirect.com/science/article/pii/S0168900213015064}
  {\bibfield  {journal} {\bibinfo  {journal} {Nuclear Instruments and Methods
  in Physics Research A}\ }\textbf {\bibinfo {volume} {740}},\ \bibinfo {pages}
  {173--179} (\bibinfo {year} {2014})}\BibitemShut {NoStop}%
\bibitem [{\citenamefont {Kumar}, \citenamefont {Pukhov},\ and\ \citenamefont
  {Lotov}(2010)}]{Kumar2010}%
  \BibitemOpen
  \bibfield  {author} {\bibinfo {author} {\bibfnamefont {N.}~\bibnamefont
  {Kumar}}, \bibinfo {author} {\bibfnamefont {A.}~\bibnamefont {Pukhov}}, \
  and\ \bibinfo {author} {\bibfnamefont {K.}~\bibnamefont {Lotov}},\ }\bibfield
   {title} {\enquote {\bibinfo {title} {Self-modulation instability of a long
  proton bunch in plasmas},}\ }\href {\doibase 10.1103/PhysRevLett.104.255003}
  {\bibfield  {journal} {\bibinfo  {journal} {Physical Review Letters}\
  }\textbf {\bibinfo {volume} {104}},\ \bibinfo {pages} {255003} (\bibinfo
  {year} {2010})}\BibitemShut {NoStop}%
\bibitem [{\citenamefont {Turner}\ and\ \citenamefont {et.
  al.}(2020)}]{Marlene2019}%
  \BibitemOpen
  \bibfield  {author} {\bibinfo {author} {\bibfnamefont {M.}~\bibnamefont
  {Turner}}\ and\ \bibinfo {author} {\bibnamefont {et. al.}} (\bibinfo
  {collaboration} {AWAKE Collaboration}),\ }\bibfield  {title} {\enquote
  {\bibinfo {title} {Experimental study of wakefields driven by a
  self-modulating proton bunch in plasma},}\ }\href {\doibase
  10.1103/PhysRevAccelBeams.23.081302} {\bibfield  {journal} {\bibinfo
  {journal} {Phys. Rev. Accel. Beams}\ }\textbf {\bibinfo {volume} {23}},\
  \bibinfo {pages} {081302} (\bibinfo {year} {2020})}\BibitemShut {NoStop}%
\bibitem [{\citenamefont {Allen}\ \emph {et~al.}(2012)\citenamefont {Allen},
  \citenamefont {Yakimenko}, \citenamefont {Babzien}, \citenamefont {Fedurin},
  \citenamefont {Kusche},\ and\ \citenamefont {Muggli}}]{Allen2012}%
  \BibitemOpen
  \bibfield  {author} {\bibinfo {author} {\bibfnamefont {B.}~\bibnamefont
  {Allen}}, \bibinfo {author} {\bibfnamefont {V.}~\bibnamefont {Yakimenko}},
  \bibinfo {author} {\bibfnamefont {M.}~\bibnamefont {Babzien}}, \bibinfo
  {author} {\bibfnamefont {M.}~\bibnamefont {Fedurin}}, \bibinfo {author}
  {\bibfnamefont {K.}~\bibnamefont {Kusche}}, \ and\ \bibinfo {author}
  {\bibfnamefont {P.}~\bibnamefont {Muggli}},\ }\bibfield  {title} {\enquote
  {\bibinfo {title} {Experimental study of current filamentation
  instability},}\ }\href {\doibase 10.1103/PhysRevLett.109.185007} {\bibfield
  {journal} {\bibinfo  {journal} {Phys. Rev. Lett.}\ }\textbf {\bibinfo
  {volume} {109}},\ \bibinfo {pages} {185007} (\bibinfo {year}
  {2012})}\BibitemShut {NoStop}%
\bibitem [{\citenamefont {Pukhov}(1999)}]{pukhov_1999}%
  \BibitemOpen
  \bibfield  {author} {\bibinfo {author} {\bibfnamefont {A.}~\bibnamefont
  {Pukhov}},\ }\bibfield  {title} {\enquote {\bibinfo {title}
  {{Three-dimensional electromagnetic relativistic particle-in-cell code VLPL
  (Virtual Laser Plasma Lab)}},}\ }\href {\doibase 10.1017/S0022377899007515}
  {\bibfield  {journal} {\bibinfo  {journal} {J. Plasma Phys.}\ }\textbf
  {\bibinfo {volume} {61}},\ \bibinfo {pages} {425--433} (\bibinfo {year}
  {1999})}\BibitemShut {NoStop}%
\bibitem [{\citenamefont {Olsen}, \citenamefont {Adli},\ and\ \citenamefont
  {Muggli}(2018)}]{Olsen2018}%
  \BibitemOpen
  \bibfield  {author} {\bibinfo {author} {\bibfnamefont {V.~K.~B.}\
  \bibnamefont {Olsen}}, \bibinfo {author} {\bibfnamefont {E.}~\bibnamefont
  {Adli}}, \ and\ \bibinfo {author} {\bibfnamefont {P.}~\bibnamefont
  {Muggli}},\ }\bibfield  {title} {\enquote {\bibinfo {title} {Emittance
  preservation of an electron beam in a loaded quasilinear plasma wakefield},}\
  }\href {\doibase 10.1103/PhysRevAccelBeams.21.011301} {\bibfield  {journal}
  {\bibinfo  {journal} {Phys. Rev. Accel. Beams}\ }\textbf {\bibinfo {volume}
  {21}},\ \bibinfo {pages} {011301} (\bibinfo {year} {2018})}\BibitemShut
  {NoStop}%
\bibitem [{\citenamefont {Lu}\ \emph {et~al.}(2005)\citenamefont {Lu},
  \citenamefont {Huang}, \citenamefont {Zhou}, \citenamefont {Mori},\ and\
  \citenamefont {Katsouleas}}]{Lu2005}%
  \BibitemOpen
  \bibfield  {author} {\bibinfo {author} {\bibfnamefont {W.}~\bibnamefont
  {Lu}}, \bibinfo {author} {\bibfnamefont {C.}~\bibnamefont {Huang}}, \bibinfo
  {author} {\bibfnamefont {M.~M.}\ \bibnamefont {Zhou}}, \bibinfo {author}
  {\bibfnamefont {W.~B.}\ \bibnamefont {Mori}}, \ and\ \bibinfo {author}
  {\bibfnamefont {T.}~\bibnamefont {Katsouleas}},\ }\bibfield  {title}
  {\enquote {\bibinfo {title} {Limits of linear plasma wakefield theory for
  electron or positron beams},}\ }\href {\doibase 10.1063/1.1905587} {\bibfield
   {journal} {\bibinfo  {journal} {Physics of Plasmas}\ }\textbf {\bibinfo
  {volume} {12}},\ \bibinfo {pages} {063101} (\bibinfo {year}
  {2005})}\BibitemShut {NoStop}%
\bibitem [{\citenamefont {Lotov}(2013)}]{Lotov2014}%
  \BibitemOpen
  \bibfield  {author} {\bibinfo {author} {\bibfnamefont {K.~V.}\ \bibnamefont
  {Lotov}},\ }\bibfield  {title} {\enquote {\bibinfo {title} {Excitation of
  two-dimensional plasma wakefields by trains of equidistant particle
  bunches},}\ }\href {\doibase 10.1063/1.4819720} {\bibfield  {journal}
  {\bibinfo  {journal} {Phys. Plasmas}\ }\textbf {\bibinfo {volume} {20}},\
  \bibinfo {pages} {083119} (\bibinfo {year} {2013})}\BibitemShut {NoStop}%
\bibitem [{\citenamefont {Lotov}(2020)}]{Lotov2020}%
  \BibitemOpen
  \bibfield  {author} {\bibinfo {author} {\bibfnamefont {V.}~\bibnamefont
  {Lotov}, \bibfnamefont {K.V.;~Minakov}},\ }\bibfield  {title} {\enquote
  {\bibinfo {title} {Proton beam self-modulation seeded by electron bunch in
  plasma with density ramp},}\ }\href {\doibase 10.1088/1361-6587/abba42}
  {\bibfield  {journal} {\bibinfo  {journal} {Plasma Phys. Control. Fusion}\
  }\textbf {\bibinfo {volume} {62}},\ \bibinfo {pages} {115025} (\bibinfo
  {year} {2020})}\BibitemShut {NoStop}%
\bibitem [{\citenamefont {Rosenzweig}\ \emph {et~al.}(2010)\citenamefont
  {Rosenzweig}, \citenamefont {Andonian}, \citenamefont {Ferrario},
  \citenamefont {Muggli}, \citenamefont {Williams}, \citenamefont {Yakimenko},\
  and\ \citenamefont {Xuan}}]{Rosenzweig}%
  \BibitemOpen
  \bibfield  {author} {\bibinfo {author} {\bibfnamefont {J.~B.}\ \bibnamefont
  {Rosenzweig}}, \bibinfo {author} {\bibfnamefont {G.}~\bibnamefont
  {Andonian}}, \bibinfo {author} {\bibfnamefont {M.}~\bibnamefont {Ferrario}},
  \bibinfo {author} {\bibfnamefont {P.}~\bibnamefont {Muggli}}, \bibinfo
  {author} {\bibfnamefont {O.}~\bibnamefont {Williams}}, \bibinfo {author}
  {\bibfnamefont {V.}~\bibnamefont {Yakimenko}}, \ and\ \bibinfo {author}
  {\bibfnamefont {K.}~\bibnamefont {Xuan}},\ }\bibfield  {title} {\enquote
  {\bibinfo {title} {Plasma wakefields in the quasi‐nonlinear regime},}\
  }\href {\doibase 10.1063/1.3520373} {\bibfield  {journal} {\bibinfo
  {journal} {AIP Conference Proceedings}\ }\textbf {\bibinfo {volume} {1299}},\
  \bibinfo {pages} {500--504} (\bibinfo {year} {2010})}\BibitemShut {NoStop}%
\end{thebibliography}%

\end{document}